\documentclass{kluwer}    

\newdisplay{guess}{Conjecture}
\usepackage[]{epsfig}
\begin{document}                                                                                   
\begin{article}
\begin{opening}         
\title{Potential of the Surface Brightness Fluctuation method to measure
            distances to dwarf ellipticals in nearby clusters} 
\author{Steffen \surname{Mieske}}  
\institute{Departamento de Astronom\'{\i}a de la Pontificia Universidad
            Cat\'{o}lica de Chile (PUC), Sternwarte der
            Universit\"at Bonn}
\author{Michael \surname{Hilker}}
\institute{Sternwarte der Universit\"at Bonn}
\author{Leopoldo \surname{Infante}}
\institute{Departamento de Astronom\'{\i}a de la PUC}
\runningauthor{Steffen Mieske et al.}
\runningtitle{Distances to dEs with the SBF Method}
\date{November 29, 2002}

\begin{abstract}
The potential of the Surface Brightness Fluctuation (SBF) method to determine
distances to dwarf ellipticals in nearby clusters is investigated. We find
that for the Fornax cluster,
the SBF method yields reliable results down to very faint magnitudes, in our
case down to $M_V\simeq-11$ mag when observing for about 1 hour at good seeing with an
8m-class telescope in the $I$-filter. Comparison between real and simulated
data for the Centaurus Cluster shows that our simulations do not
overestimate the achievable S/N of the SBF method. 
\end{abstract}

\end{opening}        

\section{Introduction}  
In this contribution, we investigate the potential of the surface brightness
fluctuation (SBF) method (Tonry \& Schneider \cite{Tonry88}) to unambigously determine cluster membership of large
samples of  dwarf ellipticals (dEs) in nearby clusters. The SBF method exploits 
the fact that on a
galaxy image each seeing disc contains a finite number of stars. This
number is quadratically proportional to distance and thus its relative rms
(causing the SBF) is inversely proportional to distance.\\
In the context of the SBF method, the distance modulus of a galaxy is given by the difference between apparent and absolute fluctuation magnitude
$(\overline{m}_I-\overline{M}_I)$. $\overline{m}_I$ is measured directly, while $\overline{M}_I$ is derived from a distance independent observable, mostly colour.\\
Tonry et al. \cite{Tonry97} have determined a well defined relation between $(V-I)$ and $\overline{M}_I$ for nearby giant early type galaxies in the range $1.0<(V-I)<1.3$:
\begin{equation}
\overline{M}_I=-1.74 + 4.5 \times ((V-I) - 1.15)\;{\rm mag}
\label{sbfrel}
\end{equation}
Regarding dwarf galaxies, the SBF method has only been applied to small samples of nearby
dEs (e.g. Jerjen et al. \cite{Jerjen01}). With the arrival of wide field imagers on
large telescopes, the use of SBF to measure distances to large sets of 
dEs in nearby clusters will be possible.
\section{Simulating and measuring SBF of dEs at the Fornax cluster distance}
The simulations were performed by adding artificial galaxy images with
implemented SBF signal onto a background field image obtained with VLT and
FORS1 in the I-filter. The exposure time was 3000 sec, the seeing 0.5$''$,
and the pixel scale 0.2$''$/pixel. The completeness limit for point sources was
I $\simeq24.8$ mag. The colour range investigated was $0.6<(V-I)<1.3$
mag. $\overline{M}_I [(V-I)]$ was derived from Tonry et al.'s
\cite{Tonry97} equation (1) and
for $(V-I)<1.0$ from Worthey's \cite{Worthe94} stellar evolutionary models
for old and intermediate age stellar populations. A distance modulus of 31.4
mag to Fornax (Ferrarese et al \cite{Ferrar00}) was adopted. The magnitude-surface
brightness relation was chosen to fit the relation observed for Fornax dEs
by Hilker et al. \cite{Hilker03}. An exponential intensity profile was adopted.

To measure the SBF, first an elliptical model of the galaxy light was subtracted
from the image. The resulting image was divided by the square root of the
model. On this image, contaminating point and extended sources were masked
out. The contribution of contaminating sources below the detection limit was
negligible.

Of the cleaned image, the power spectrum (PS) was calculated and azimuthally
averaged. The result can be written in the form:
\begin{equation}
P(k)=PSF(k)\times P_0 + P_1
\end{equation}
$PSF(k)$ is the PS of the seeing profile. $P_0$ is the amplitude of the SBF,
given in ADU. $P_1$ is the white noise component. Fig.~1 shows an example
image and power spectrum for a simulated galaxy.
\begin{figure*}
\hspace{0.8cm}
\epsfig{figure=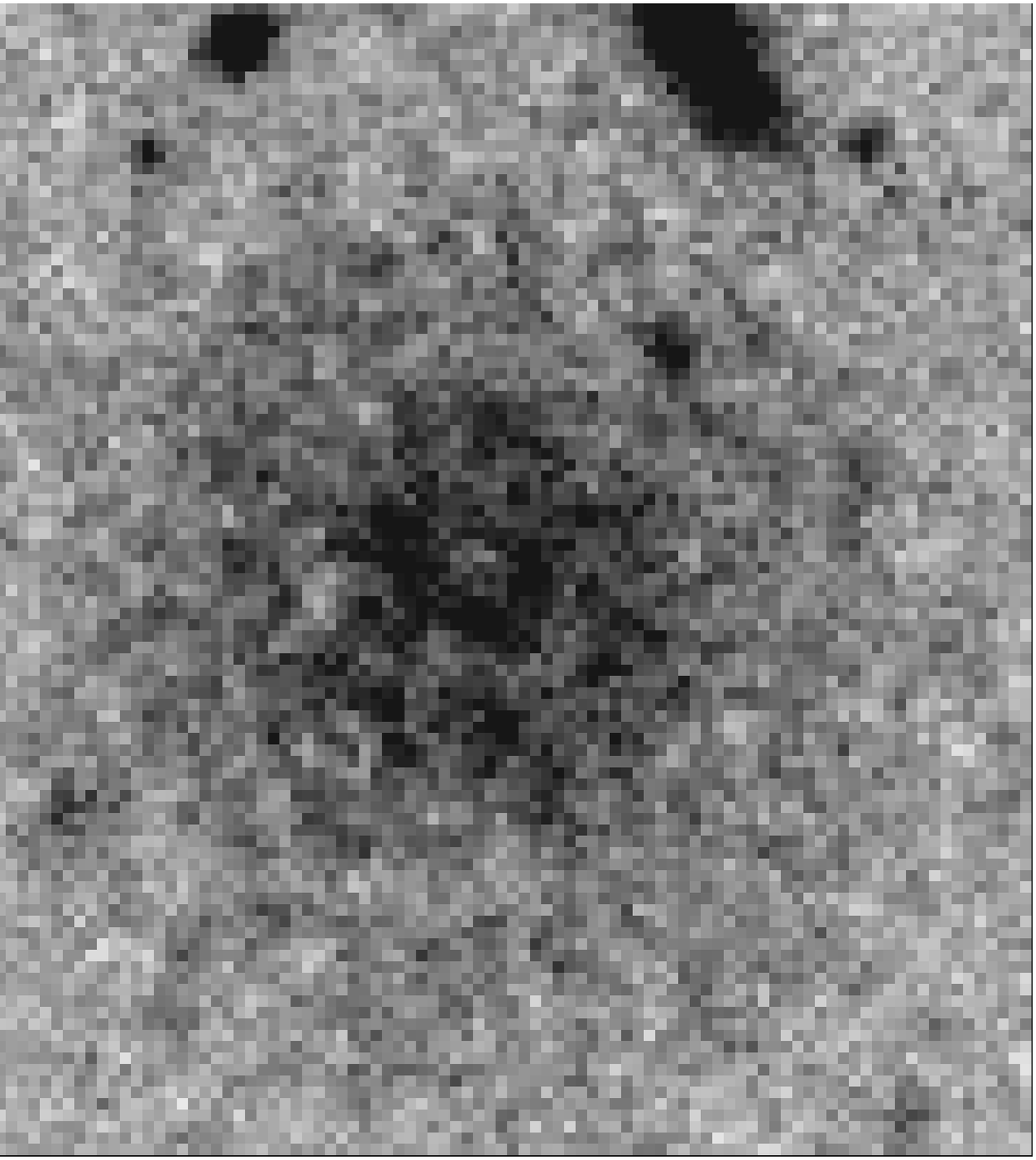,width=3.3cm}\hspace{0.4cm}
\epsfig{figure=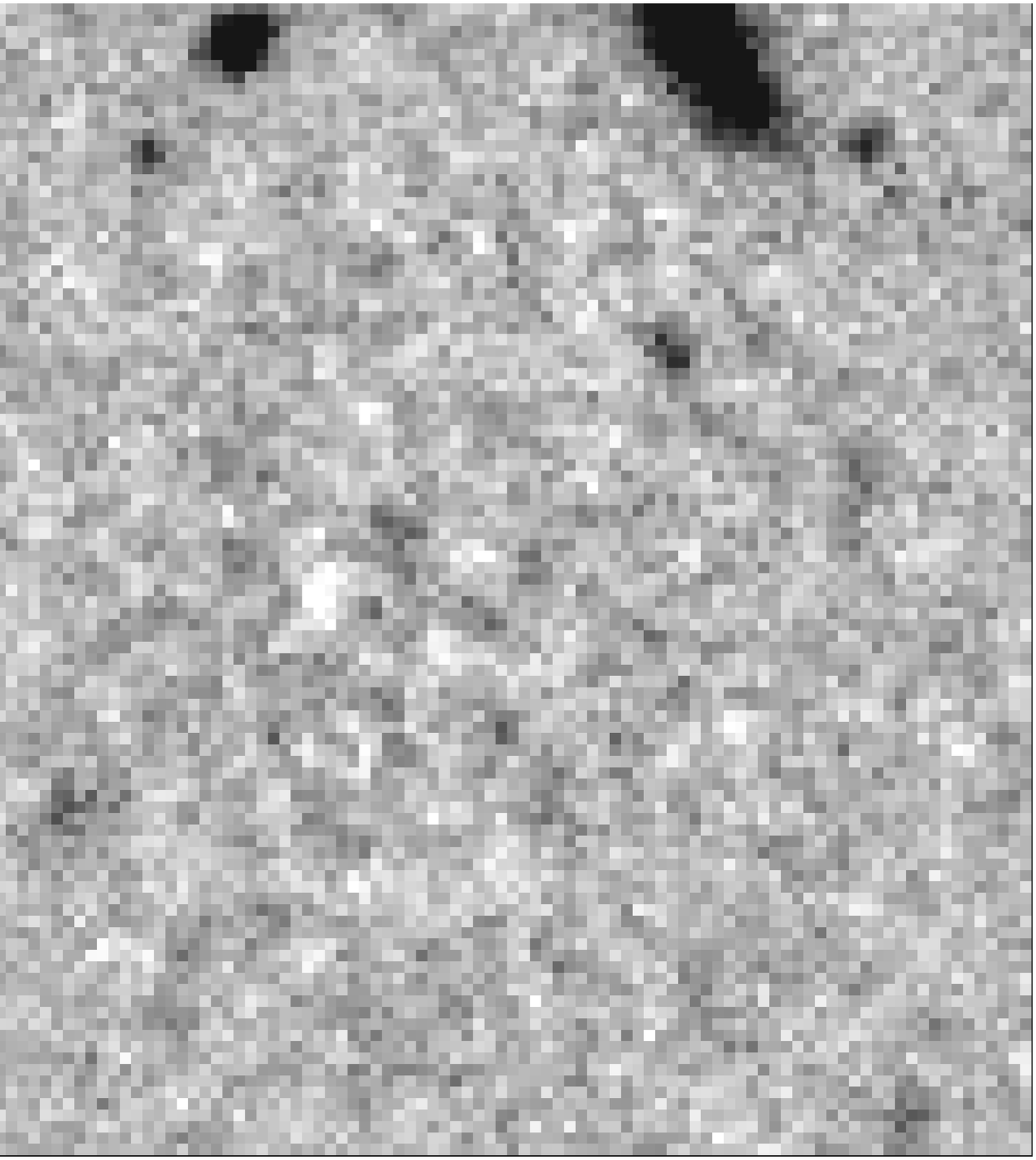,width=3.3cm}
\epsfig{figure=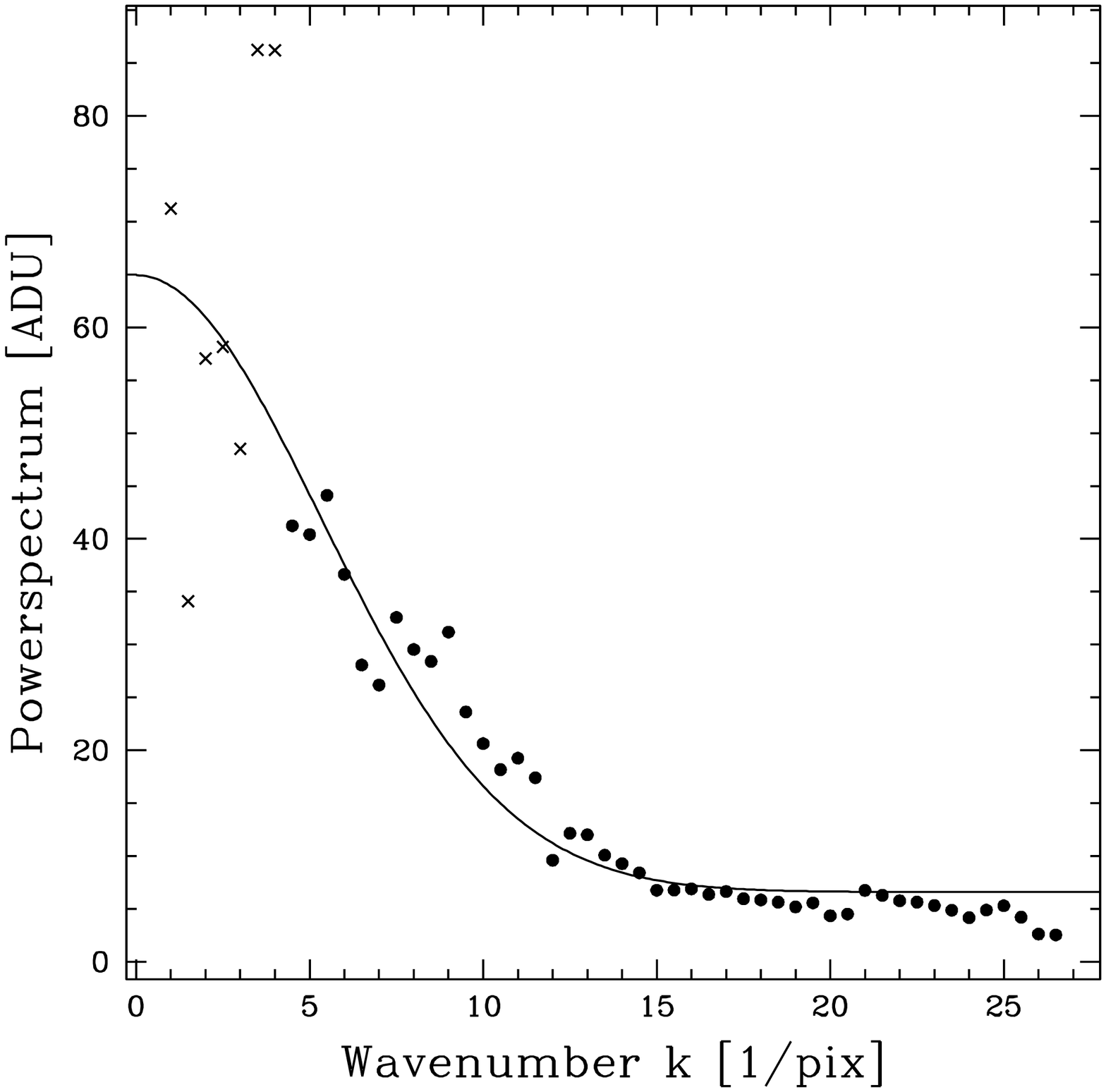,width=3.6cm}\\
\caption{\label{measproc}
Typical example for a simulated  dE in the I band before (left panel) and after
(middle panel) subtraction of an elliptical galaxy model. The right panel shows the
 azimuthally averaged power spectrum of the SBF fluctuations. The solid line is the
best fit of $P(k)=PSF(k)\times P_{0}+P_1$, when rejecting the points marked with 
crosses. The measured apparent SBF magnitude is correct within 0.15 mag.
Parameters of the simulated dE: V=18.7 ($M_V=-12.6$) mag, $\mu_V$=24.35 mag/arcsec$^2$,
$(V-I)$=0.87, $\bar{M}_I$=-2.92 mag.}
\end{figure*}
\section{Results of the simulations}
Fig.~2 summarizes the results of the SBF measurements on the simulated
dEs. For $M_V=-11$ mag, about 50\% of the measured galaxies have SBF with
S/N$>3$ and
$\delta\bar{m}_I=\bar{m}_{I,\rm simulated}-\bar{m}_{I,\rm measured}$ smaller than
0.5 mag. As nearby clusters such as Fornax or Virgo are isolated from background galaxies by several magnitudes in distance modulus, the SBF method therefore allows for unambigous cluster membership determination even for low S/N SBF data.
\begin{figure}
\begin{center}
\psfig{figure=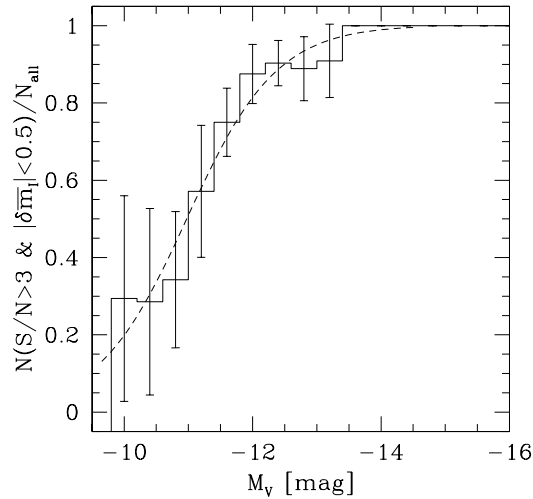,width=7cm}
\end{center}
\vspace{-0.2cm}
\caption[]{\label{resultsfig}Results of the simulations: 
Solid histogram: fraction of SBF measurements with S/N$>$3 and measurement
accuracy $|\delta\bar{m}_I| < 0.5$ mag plotted vs. $M_V$. Dashed curve: Fit
to the data fainter than $M_V=-12.8$ mag. The 50\% limit is reached at about
$M_V\simeq-11$ mag.}
\end{figure}
\subsection{Simulations vs. Reality}
To demonstrate that our simulations do not overestimate the S/N of the SBF
measurement on real galaxies, we compared SBF measurements for Centaurus
Cluster galaxies, obtained from VLT-FORS1 images in the I-filter, with
simulations tuned to reproduce the measured values for these galaxies. As
one can see in Fig.~3, the S/N of the simulations matches well the S/N of
real galaxy measurements.
\begin{figure}
\begin{center}
\psfig{figure=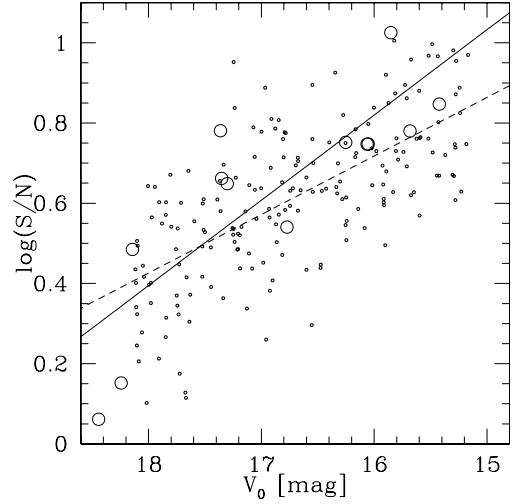,width=7cm}
\end{center}
\vspace{-0.2cm}
\caption[]{\label{sbfcensim}Comparison between simulations and reality:
S/N of the SBF measurement vs. apparent $V$ magnitude of the galaxy.
Small circles represent simulated galaxies, with $\bar{m}_I$ according to
the SBF-colour-magnitude relation found for the real data. Large circles
are the real data. The solid line represents a fit to the real data,
the dashed line is a fit to the simulated data.}
\end{figure}
\section{Conclusions}
It has been shown that for nearby clusters such as the Fornax cluster, the
SBF Method can yield reliable membership determination for dEs down to
very faint magnitudes. We find that our simulations do not overestimate the
achievable S/N, but are in good agreement with real measurements.
An ideal application of the SBF method would be a deep wide field
survey of nearby clusters such as Fornax or Virgo.

\theendnotes

\end{article}
\end{document}